\documentclass[journal]{IEEEtran}
\usepackage[pdftex]{graphicx}

\usepackage{cite}

\DeclareGraphicsExtensions{.pdf,.jpeg,.png}

\usepackage{amsmath,amsfonts,amssymb}
\usepackage{gensymb}

\usepackage{paralist}
\usepackage{textcomp}

\makeatletter
\let\MYcaption\@makecaption
\makeatother

\makeatletter
\let\@makecaption\MYcaption
\makeatother

\usepackage{url}
\usepackage{hyperref}

\newcommand{\fig}[1]{Fig.~\ref{fig:#1}}

\begin{document}
\bstctlcite{IEEEexample:BSTcontrol}
%
\title{A Multi-Core Fibre Photonic Lantern Based Spectrograph for Raman Spectroscopy}

%
%
%

\author{Christopher~H.~Betters,
Joss~Bland-Hawthorn,
Salah~Sukkarieh,
Itandehui~Gris-Sanchez,
and~Sergio~G.~Leon-Saval,~\IEEEmembership{Member,~IEEE}

\thanks{Manuscript received January~24,~2020; revised February~24,~2020; accepted February~24,~2020}
\thanks{C. H. Betters, J. Bland-Hawthorn and S. G. Leon-Saval are with the Sydney Astrophotonic Instrumentation Laboratory and Sydney Institute for Astronomy, School of Physics, The University of Sydney, NSW, Australia
 e-mail: (christopher.betters@sydney.edu.au).}
\thanks{S. Sukkarieh is with the Australian Centre for Field Robotics, The University of Sydney, NSW, Australia}
\thanks{I. Gris-Sanchez is with the Department of Physics, University of Bath, Claverton Down, Bath BA2 7AY, UK currently at ITEAM Research Institute, Universitat Polit\`ecnica de Val\`encia,Valencia, 46022, Spain}
\thanks{Thanks to Prof. Tim A. Birks from the University of Bath for  facilitating the fibre fabrication and the use of the fibre drawing tower}
\thanks{This work was supported in part by the University of Sydney SREI 2020 and in part by JBH's ARC Laureate Fellowship (FL140100278).}
\vspace*{-2.5\baselineskip}}

%
%

\markboth{IEEE PHOTONICS TECHNOLOGY LETTERS, VOL. TBD, NO. TBD, TBD}%
{Shell \MakeLowercase{\textit{et al.}}: Bare Demo of IEEEtran.cls for IEEE Journals}
%



\maketitle
\setcounter{page}{1}

\begin{abstract}
We report on the development of a compact (volume $\approx$ 100\:cm$^3$), multimode diffraction-limited Raman spectrograph and probe designed to be compact as possible.
The spectrograph uses  `off the shelf' optics, a custom  3D-printed two-part housing and harnesses a multi-core fibre (MCF) photonic lantern (multimode to few-mode converter), which slices a large 40~\textmu m multimode input into a near-diffraction-limited 6~\textmu m aperture.
Our unique design utilises the hexagonal geometry of our MCF, permitting high multimode collection efficiency with near-diffraction-limited performance in a compact design.
Our approach does not require a complex reformatter or mask and thus preserves spectral information and throughput when forming the entrance slit of the spectrograph. 
We demonstrate the technology over the interval 800~nm to 940~nm (200~cm$^{-1}$ to 2000~cm$^{-1}$) with a resolution of 0.3\:nm (4\:cm$^{-1}$), but other spectral regions and resolutions from the UV to the near infrared are also possible.
We demonstrate the performance of our system by recording the Raman spectra of several compounds, including the pharmaceuticals paracetamol and ibuprofen.
\end{abstract}

\begin{IEEEkeywords}
Raman spectroscopy; Optical spectroscopy; Photonic lantern; Multi-core fibre; 3D Printed; Fibre reformater.
\end{IEEEkeywords}

%
\IEEEpeerreviewmaketitle
\vspace*{-1.5\baselineskip}
\section{Introduction}
\IEEEPARstart{R}{aman} spectroscopy is a commonly used, non-invasive, optical analytic technique in pharmaceuticals and bioscience as it can provide excellent chemical specificity. Commercially available Raman spectrographs require significant investment, even more so if a bespoke design is needed. Here we present a portable, customisable low-cost Raman sensor enabled by advanced photonic devices and a unique 3D-printed housing.

Our design was motivated in part by the recent trend of low-cost 3D printers that has seen many researchers building their own low cost `Smartphone' spectral sensor for a variety of applications \cite{Wilkes2017,Liu2017,Hossain2016}.
Most of these have either been add-ons to augment smartphones, while others seek to harness just the smartphone-style CMOS image sensors made cheap by mass production.
There a few issues that arise with using such sensors, e.g. the embedded filters built into the detector array such as a Bayer filter or a near-infrared cutoff filter.
While these filters can be removed with due care \cite{Wilkes2016}, for precision science applications, it is more appropriate to use the monochrome version of the sensor manufactured without any colour filters such as astronomy science grade sensors.
Many of these sensors also have tiny pixels ($\sim$1~\textmu m), which will limit light collecting ability and its achievable signal to noise \cite{Farrell2006}.
Smaller pixels also require faster (low f/\#; and more complex) optics to sample the input slit image formed by a spectrograph appropriately.
Fortunately, there is a robust market for monochrome sensors, with a range of pixel sizes, thanks to various requirements in machine vision applications and those of groups like amateur astronomers (many premium sensors in this market rival 'science grade' sensors at a very reasonable cost).

To enable a compact spectrograph design that is ideally suited to these sensors, we use a device called the Photonic Lantern (PL) \cite{Leon-Saval2013,Birks2015}. 
The PL is a photonic multimode (MM) to single-mode (SM) converter, where the number of modes in the MM end is converted to the equal number of SM or few-mode outputs, thus conserving the entropy and allowing low losses in the process.
This effectively allows us to remap a large spectrograph slit into a minimum width slit (diffraction-limited or SM in the ideal case), thereby allowing an ultra-compact optical design.  We can then combine both the optimal light-collecting ability of an MMF and also the design simplification of SMFs (or cores) being used as the spectrograph input slit \cite{Bland-Hawthorn2010}.

The PL used here is made using a 19 multi-core fibre, instead of a bundle of SMFs\cite{Leon-Saval2013, Birks2015}, that is tapered with a low-index glass capillary jacket to produce a 40~\textmu m MM input.
This will take the place of the MM collection fibre in our Raman probe.
The other end of the MCF is used directly as the spectrograph slit, with the array forming a pseudo-slit. This technique is unique to our design.
This can only be done if the fibres cores are sufficiently spaced in a regular grid (hexagonal in this case), such that as when the grid is rotated, each core becomes vertically separated, and spectrally dispersed without overlapping its neighbours when imaged onto the spectrograph sensor.

\IEEEpubidadjcol

\begin{figure}[t]
    \centering
    \includegraphics[width=\linewidth]{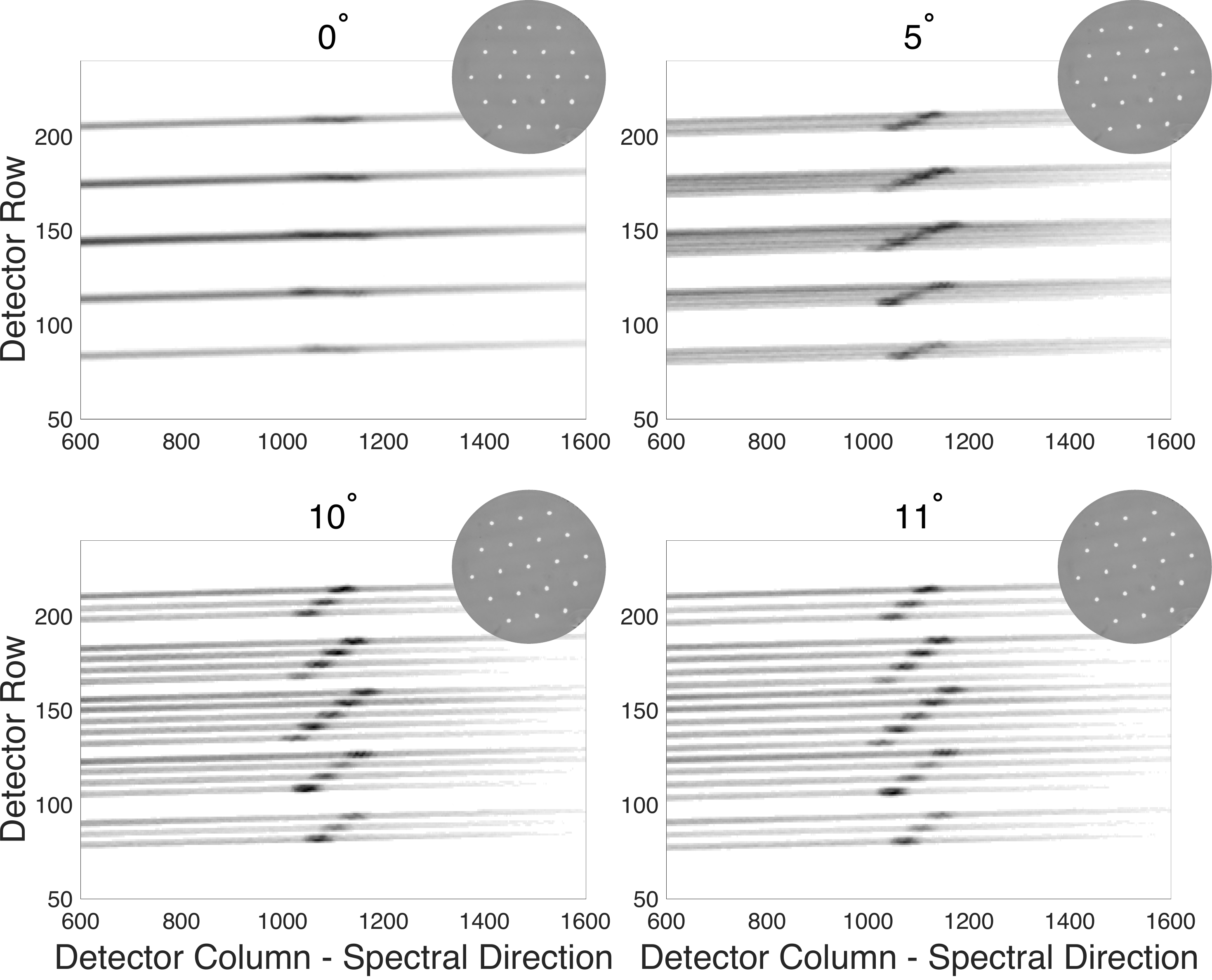}
    \caption{Spatial versus spectral coordinates detector images of the Raman spectrograph being fed by a halogen light source and a narrow band filtered 850\:nm LED source for 4 different MCF pseudo-slit angle rotations 0\degree, 5\degree, 10\degree and 11\degree. As the MCF is rotated the spectra of each individual core becomes vertically separated (and spectrally offset). At 11 \degree they become fully distinct. Note - only a portion of the whole detector image is shown here.
    }
    \label{fig:TigerRoation}
\end{figure}

We call this the `Photonic TIGER' configuration \cite{Leon-Saval2012}, named for its inspiration: the micro-lens array based TIGER spectrograph \cite{Bacon1995}. 
The principle is illustrated in \fig{TigerRoation}, where we show the resulting spectra with the MCF pseudo-slit rotated 0, 5, 10, 11 degrees.
The actual rotation angle, $\theta$, of the hexagonal grid needed to achieve this is dependant on the number of cores as follows:
\begin{equation}\label{eqn:theta}
\tan \theta  = \textstyle\frac{{\sqrt 3 }}{{2{N_{{\text{mid}}}} - 1}}
\end{equation}
where $N_\text{mid}$ is the number of cores on the central row of the hex grid. The final vertical core-to-core spacing, $v_\text{sep}$, is then given by 
\begin{equation}\label{eqn:adj_sep}
v_\text{sep}=d_\text{grid}\sin\theta.
\end{equation}
where $d_\text{grid}$ is the grid spacing.

To have sufficient vertical spacing for the TIGER configuration, and due to fibre availability, here we have used an MCF (with 6.5~\textmu m diameter cores separated by 60~\textmu m) that was originally intended for the telecom-infrared window around 1550~nm.
As a result, even though the optical design presented here supports a true SM input, we are currently operating in the `few-mode' (2-3 modes per core)  quasi-diffraction limited regime.
In practice, this produces a broadened point-spread function (and a corresponding reduction in effective spectral resolution).
This is not a fundamental issue, but a by-product of the TIGER pseudo-slit geometry and the particular preform used to manufacture our MCF.
Operating in this few-mode' regime should actually improve light collecting ability and further, it demonstrates the possibility of an optical fibre slicer to operate in a MM to few-mode architecture as proposed in \cite{Leon-Saval2017}. Nonetheless, both a visible wavelength version of the MCF and an IR version of the spectrograph are planned future investigations.

In the following sections, we discuss the design of our Raman spectrograph and its corresponding probe, that was originally designed and developed for deployment aboard autonomous vehicles in the farming industry \cite{ladybird}. To the best of our knowledge, this is the first demonstration of such a device, a combination of 3D printed Raman spectrograph with advanced photonic technologies -- a multi-core fibre Photonic Lantern.

\vspace*{-1\baselineskip}
\section{Design}
The actual spectrograph is relatively conventional in the sense that it is comprised of bulk optics  (i.e. lenses, prisms, a diffraction grating) components.
However, when we assume an SMF pseudo-slit as the input, the design can be considered genuinely diffraction-limited\cite{Betters2012,Betters2013}.
Further, it pairs better with detectors with small pixels, which are more prevalent commercially (and thus cheaper) thanks to the mass production for smartphone cameras and machine vision\cite{Bland-Hawthorn2010}.

The optical design for our Raman system was developed around a few key parameters required by our applied science application. 
For our proof of concept, we are using a 785~nm excitation laser, and our maximum measurement range will be from 785~nm to 1000~nm (i.e. a Raman scattering of 0~cm$^{-1}$ to 2700~cm$^{-1}$; practically the range is smaller due to filters to block excitation laser and detector sensitivity. 
Our goal was to have around 790~nm to 940~nm (100~cm$^{-1}$ to 2000~cm$^{-1}$) consistent with other Raman spectrographs in the market.
Further, we also wanted to maximise the spectral resolution for this wavelength range while only using commercial `off-the-shelf' parts.
The two key determining factors were:
\begin{inparaenum}[1)]
    \item the detector pixel scale and count -- in our design we use a PointGrey Grasshopper3 GS3-U3-32S4M-C a 2048 x 1536 sensor with 3.45~\textmu m pixels (\href{https://www.ptgrey.com/grasshopper3-32-mp-mono-usb3-vision-sony-pregius-imx252}{ptgrey.com}); and
    \item to a lesser extent, the grating line density -- the grating also needed to be blazed for the correct wavelength range. For simplicity, we opted for a transmissive spectrograph using a volume phase holographic (VPH) grating in the Littrow configuration. Here we chose a Wasatch Photonics Volume Phase Holographic grating with a line density of 1200~l/mm with peak efficiency at 840~nm (\href{https://wasatchphotonics.com/product/1200-lmm-at-840nm/}{wasatchphotonics.com}).
\end{inparaenum}

Assuming 1024 resolution elements determined by 2 pixels per FWHM to sample sufficiently with the PointGrey Camera, the spectral resolution would be around 0.15~nm (2.4cm$^{-1}$) with a resolving power of 6000. For comparison, commercially available `portable' Raman spectrographs tend to have spectral resolution closer to 10~$cm^{-1}$, with higher resolutions being reserved for larger bench-top models.
This spectral resolution, in combination with our 1200 l/mm grating, lets us then determine the size of the collimated beam (and thus the effective focal length of the collimator when combined with the output of the optical fibre.  Assuming the benefits of a system with a diffraction-limited slit, we get a 5mm beam (i.e. $R=mN$).
We use this information as a starting point in Zemax OpticStudio (\href{https://www.zemax.com/opticstudio}{zemax.com}) and then substitute appropriate lenses.
The final design uses a 15mm and 30mm achromatic doublets to produce a collimator with an effective focal length (EFL) of 16mm.
To correctly sample the PSF with the detector we need a magnification of around a factor of two (as the FWHM mode field diameter of the MCF is similar in size to the PointGrey detector pixels). 
This sampling is achieved using two 60mm achromatic doublets to form a camera with an EFL of 32mm.
All the lenses used are stock Thorlabs (\href{https://www.thorlabs.com}{thorlabs.com}) components. 
The compactness of this design means that the entire optical design can comfortably fit within the print envelope of many 3D-printers, allowing us to implement a complex custom housing that would be impractical to conventionally machine.

\begin{figure}[t]
    \centering
    \includegraphics[width=0.9\linewidth]{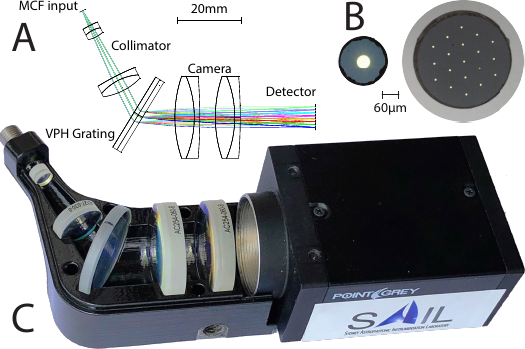}
    \caption{
    {\bf(a)} The OpticStudio ray-trace showing the transmissive spectrograph design.
    {\bf(b)} Microscope image of the MCF and PL end-face's at the same scale (both are illuminated with white light to clearly show fibre cores).
    {\bf(c)} Image showing the bottom half of our 3D-spec housing mounted to a PointGrey Grasshopper3 sensor and with optics in-place. The only alignment requirements are rotation of the diffraction grating, rotation of the MCF input and finally focus of the camera of the sensor. 
    }
    \label{fig:3dspec}
\end{figure}

We commence our housing design by exporting a full 3D model of the optical design from OpticStudio using the IGES format and importing into Solidworks 3D-cad (\href{https://www.solidworks.com}{solidworks.com}).
The housing model is built around the optical design, starting with a solid block encompassing the optics, and then removing appropriate gaps to hold the optics, provide a beam path, mount a fibre connector and attach the detector.
An advantage of the particular printer we use, the Stratasys Objet30 Pro (\href{https://www.stratasys.com/3d-printers/objet30-pro}{stratasys.com}) is that it has sufficient resolution to directly print the 1/4-36 thread that an SMA fibre connector requires.
We use this ability to incorporate a standard SMA fibre bulkhead adapter, which helps ensure the fibre input is concentric with the collimating lenses.
The resulting housing, along with the optical design layout, can be seen in \fig{3dspec}.

This approach to the housing design eliminates many degrees of freedom from the optical alignment.
The remaining factors are the rotation of the diffraction grating (which a square grating would negate), the focus of the camera lens on the detector (achieved by moving the entire PointGrey detector along slots incorporated in the housing base) and rotation of the input fibre (to enable the TIGER dispersion mode).
The relative positions of optics and grating are all fixed, and they are all just inserted in place without the need for extra optical alignment. This type of design is a very promising approach for any type of compact spectrograph design or other optical systems.

\begin{figure}[b]
    \centering
    \includegraphics[width=0.9\linewidth]{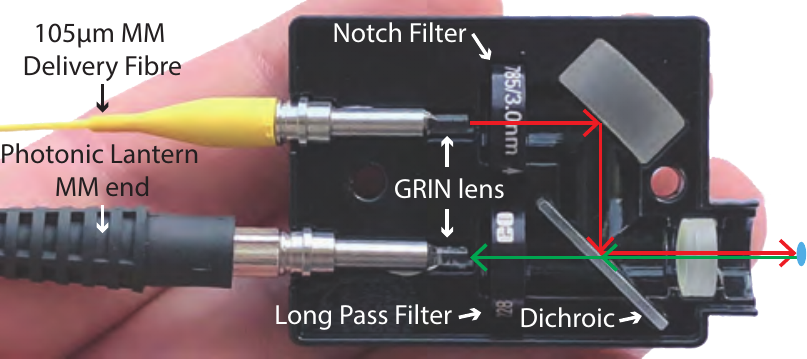}
    \caption{
    Shown is our bespoke Raman probe's 3D-printed housing without its cover and with optics in place.
    The optical path is illustrated in the left image. The red path shows the excitation laser, which is brought to a focus on the target sample (shown in blue). The green path shows the light that is collected and fed to the Raman spectrograph via the PL MM end.
    }\label{fig:probe}
\end{figure}

To incorporate our PL technology, we also built a custom Raman probe using optical filters to reject the Rayleigh scattered light and suppress background light associated with the excitation laser delivery fibre (which also has a unique fluorescence and Raman signature).
The primary goal of this probe was to suppress unwanted background signal and to maximise the collection of the Raman signal using our PL.
To accomplish this, we used three filters:
\begin{inparaenum}[a)]
  \item A narrow bandwidth notch filter, also known as a laser-line cleanup filter. This filter only allows a narrow bandwidth centred on the excitation laser wavelength to be transmitted, thereby removing any Raman and fluorescence signal generated by the delivery fibre itself.
  \item A dichroic mirror/filter which reflects the excitation laser and transmits light with wavelengths red-ward of 800~nm. It is also used to combine the light paths of the collection fibre and laser delivery fibre.
  \item A long-pass filter. We use an off-the-shelf filter from Edmund Optics (part 86228 - \href{https://www.edmundoptics.com.au/optics/optical-filters/longpass-edge-filters/785nm-12.5mm-diameter-raman-edge-filter/}{edmundoptics.com.au}) that has a cut-on wavelength of 792.8nm. This component is the primary method of suppressing the excitation laser light. 
\end{inparaenum}

In our optical design, the laser delivery fibre and PL collection fibre have been connectorised into two modified SMA connectors (we removed the threaded barrel leaving only the ferrule).  Both of these are collimated using an 850nm GRIN lens affixed to the front of the connector using UV-curing optical glue. The collimated beams from the two connectors are passed through their respective filters and are combined using the dichroic. Finally, a 12.5mm lens is used to focus/collect the light on the probe-target.

We developed the probe housing using a similar process used in the spectrograph housing design; we import a 3D model of the optical layout of the probe into Solidworks, generate a solid model encompassing all components and removed appropriate gaps for the optics and components.
The profile of the SMA connectors is also removed from the model, allowing them to be clamped into place. This helps ensure the connectors are parallel and correctly aligned with the other components.
The final layout of the optics mounted in their 3D printed housing (and the light paths) is shown in \fig{probe}.

\vspace*{-0.5\baselineskip}
\section{Results}
\begin{figure}
    \setlength{\textfloatsep}{-0.1\baselineskip plus 0.2\baselineskip minus 0.1\baselineskip}
    \centering
    \includegraphics[width=\linewidth]{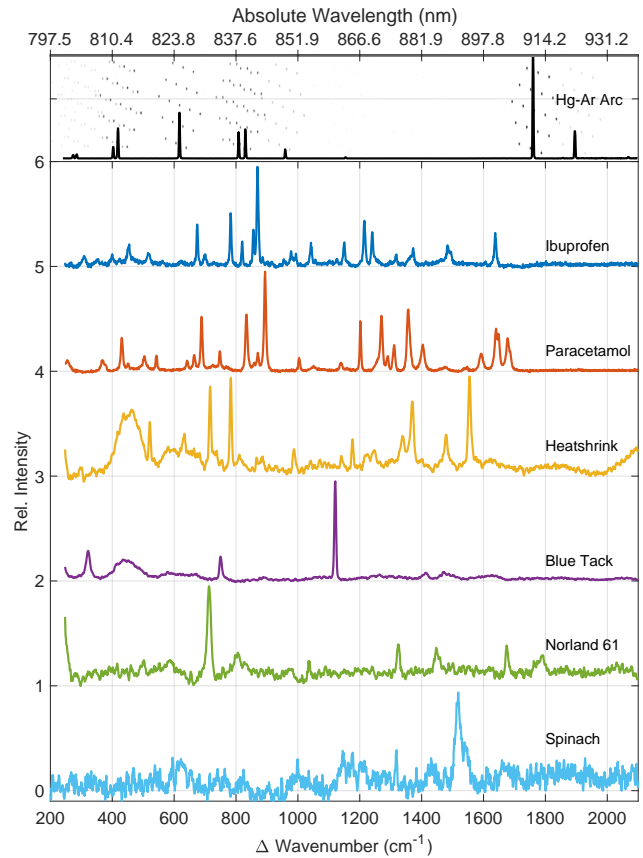}
    \caption{Shown are six example Raman spectra and a calibration arc spectrum. Some spectra have broad fluorescence features subtracted (high pass filter). In order from top to bottom these are:
    \textit{1)} Hg-Ar calibration lamp;
    \textit{2)} Ibuprofen;
    \textit{3)} Paracetamol (Acetaminophen);
    \textit{4)} Heat-shrink (polyolefin);
    \textit{5)} Blu-Tack (synthetic rubber);
    \textit{6)} Norland 61 UV adhesive.
    \textit{7)} Spinach Leaf.
    The Hg-Ar spectrum is overplotted on the full 2D image from the spectrograph detector, showing the signature of the hexagonal MCF input slit.
    }
    \label{fig:spectra}
\end{figure}

The MCF TIGER input slit means we measure 19 individual spectra simultaneously on the detector.
These are extracted using an optimal extraction algorithm that simultaneously fits the intensity in each fibre. 
This process minimises fibre-to-fibre cross-talk and limits the read-noise contribution of individual pixels. 
A mercury argon lamp (Ocean Optics HG-1 \href{https://oceanoptics.com/product/hg-1/}{oceanoptics.com}) is used to calibrate the wavelength scale.
The 19 individual spectra are then combined on a common linearised wavelength scale to produce a single 1D spectrum. 
Several example spectra, including the HgAr calibration spectrum, can be seen in \fig{spectra}. 
Some of these have been further normalised to remove broad fluorescence features of the compounds being probed.
The top panel of \fig{spectra} shows the final combined spectrum of the Hg-Ar calibration lamp overplotted on the 2D detector image.
In the plot, the Hg-Ar spectrum is aligned with the central core of the MCF, and we can easily see each line of the calibration spectrum correspond with a unique set of spots (arranged in a rotated hexagonal grid) as described in Section 2 and \fig{TigerRoation}.

We also use the Hg-Ar spectrum to measure the actual resolution of the spectrograph, giving R $\approx$ 3200 @ 843.1nm and R $\approx$ 3500 @ 914.5nm. This about half the maximum possible, due to the use of the IR MCF fibre, as explained in Section 2. Nonetheless, our operating resolution is several times higher than standard commercial portable Raman spectrographs.

We measured the throughput of the PL, with the GRIN lens attached to the multimode lantern end, used in the probe to be 68\%  by injecting light from a SMF using another 850~nm GRIN lens. The throughput was obtained by measuring the output of the SMF before the additional GRIN lens and at the output of the multi-core fibre with a large area photo-detector. This approach should measure an efficiency similar to the collection efficiency of the probe. 
To benchmark the coupling efficiency, the same approach measured a throughput of 87\% when using the 105~\textmu m Raman laser delivery fibre with the same GRIN lens attached (as shown in \fig{probe}). This gives us an upper limit and worst case scenario for the transmission loss of the lantern of 19\% (0.91~dB). Furthermore, by using the same input/output power measurement, we obtained a 90\% transmission from one of the 19 cores of the multi-core fibre to the collimated output of the PL with the GRIN lens attached.    

\vspace*{-0.5\baselineskip}
\section{Conclusion}
Our compact Raman spectrograph and probe are the result of a novel combination of advanced photonic transitions, such as the photonic lantern, and bespoke 3D-printable design. 
The PL incorporated in the probe allows use to reformat a MM collection fibre to smaller near-diffraction limited spectrograph slit, and thus enables the minimum size optical configuration (for a given detector/diffraction grating configuration).
The use of relatively low cost and low noise commercial machine vision detectors complement the reduced size requirements.
Here we have discussed the actual design process of both the probe and spectrograph, along with the unique hexagonal pseudo-slit and unique data reduction approach, that allows the MCF to feed the spectrograph directly without further optical manipulation | the Photonic TIGER mode.
Several examples of Raman spectra were given to demonstrate the versatility of the device, with clear signatures of multiple pharmaceuticals and other compounds detected with short integrations.
We would also like to note -- while the design in the paper was focused on Raman spectroscopy using a 785nm excitation source, the design process can easily be generalised to any application that benefits from or traditional use an MM fibre input.
Other similar uses we are exploring include a small testbed spectrograph aboard the i-INSPIRE 2 satellite in low Earth orbit \cite{Cairns2017} and the development of stable radial velocity spectrographs \cite{Betters2016}.






\bibliographystyle{IEEEtran}

%
\vspace*{-1\baselineskip}
\bibliography{IEEEabrv,references}

\end{document}